\documentclass[prd,preprint,floatfix,nofootinbib]{revtex4}
\usepackage{epsfig}
\usepackage{graphicx}
\usepackage{pstricks}

\newcommand{\newc}{\newcommand}
\newc{\beq}{\begin{equation}}
\newc{\eeq}{\end{equation}}
\newc{\eeqa}{\end{eqnarray}}
\newc{\beqa}{\begin{eqnarray}}
\newc{\bsym}{\boldsymbol}
\newc{\mrm}{\mathrm}
\newc{\ovl}{\overline}
\newc{\ovla}{\overleftarrow}
\newc{\ovra}{\overrightarrow}
\newc{\ra}{\rightarrow}
\newc{\lra}{\leftrightarrow}
\newc{\wtil}{\widetilde}
\newc{\eps}{\epsilon}
\newc{\hc}{\dagger}
\newc{\pd}{\partial}
\newc{\SL}{\!\!\!/}
\newc{\LH}{\hat{L}}
\newc{\RH}{\hat{R}}
\newc{\sWsq}{\sin^2\theta_\mathrm{W}}
\newc{\cWsq}{\cos^2\theta_\mathrm{W}}
\newc{\half}{\frac{1}{2}}
\newc{\hh}{\hat{H}}
\newc{\hphi}{\hat{\Phi}}
\newc{\nonr}{\nonumber}
\newc{\gev}{~\rm{GeV}}

\begin{document}

\baselineskip=15pt

\preprint{}

\title{Fourth Generations with an Inert Doublet Higgs}

\author{Efunwande Osoba}
\affiliation{Harish Chandra Research Institute, Allahabad, India}
\email{eosoba@mri.ernet.in}
\date{\today}

\vskip 1cm

\begin{abstract}
We explore an extension of the fourth generation model with multi-Higgs doublets and three fermion singlets.  The Standard Model neutrinos acquire mass radiatively at one loop level while the fourth generation neutrinos acquire a heavy tree-level mass. The model also contains several Dark Matter candidate whose stability is guaranteed by a $Z_2$ discrete symmetry. The possibility of CP violation in the scalar sector is also briefly discussed.
\end{abstract}

\maketitle

\section{Introduction}
To date, the Standard Model (SM) has proven to be a very predictive and successful model of particle physics. However, there is no reason why the SM should only have three generations, and therefore examining the SM extended with additional generations is a worthwhile effort.

Phenomenological studies of the fourth generations quarks abound in the literature. The ATLAS collaboration \cite{Aad:2012bb, ATLAS:2012aw}  excludes the bottom-like fourth generation quark $b'$  with $m_{b'} < 450 \gev$, and the top-like quark  $t'$ of $m_{t'} < 480\gev$.

The fourth generation quarks enhance the Higg-gluon-gluon fusion rate via loop contributions, and therefore the Higgs production cross section is increased by a factor of nine.  Given such an enhancement, the signature of fourth generation quarks would be unmistakable at the Tevatron and the Large Hadron Collider (LHC). The CMS collaboration excludes a SM Higgs with mass $120\gev < m_H < 600\gev$ at 95\% C.L. Atlas excludes the range $140\gev < m_H < 185\gev$ \cite{Aad:2011qi}.
 
The invisible Z-boson width, as measured by LEP, excludes fourth generation leptons less than half of the mass of the Z boson. In addition, LEP II places addition bounds, $ m_{l'} > 100 \gev$ \cite{Nakamura:2010zzi}. If the fourth generation neutrino mixes significantly with the SM neutrinos then $ m_{\nu'} > 80-100 \gev$ \cite{Carpenter:2010dt}. 

As it can be clearly seen, the fourth generation neutrino is expected to be much heavier than the SM neutrinos. The extreme hierarchy between the SM neutrinos and the fourth generation neutrino is rather puzzling.  A split mass generation mechanism can explain the hierarchy in neutrino masses. The SM neutrinos could acquire mass through radiative corrections while the fourth generation neutrino acquires a tree level Dirac and Majorana mass contributions through its Yukawa coupling with a fermion singlet, as shown in \cite{Aparici:2011nu}.  However, a five generation model was necessary in order to obtain the $\delta m^2$ as proven by the solar and atmospheric neutrino experiments.

In the light of the existing tensions in the fourth generation models, we propose another way to resolve the neutrino mass hierarchy and thus turn our attentions to multi-Higgs models with an inert doublet.    

\section{Four generations and two Higgs doublet models with an inert doublet}
Here, we shall explore extending four generations with two Higgs doublets, of which one is an inert doublet. The SM is extended with two Higgs doublets, $\Phi_1, \Phi_2$, three extra gauge singlet fermions $N_i$, an extra lepton generation  $L^4_L, L^4_R$, and quarks $U_R, U_L$. A $Z_2$ discrete symmetry is imposed to ensure that there are no flavor changing neutral current in the SM quark sector.

The scalar potential is 
\beqa
V_{12}(\Phi_1,\Phi_2) &=&-\frac12\left\{m_{11}^2\Phi_1^\dagger\Phi_1+ m_{22}^2\Phi_2^\dagger\Phi_2\right\} \nonumber \\
                                        &+&\frac{\lambda_1}{2}(\Phi_1^\dagger\Phi_1)^2+ \frac{\lambda_2}{2}(\Phi_2^\dagger\Phi_2)^2+ \lambda_3(\Phi_1^\dagger\Phi_1)(\Phi_2^\dagger\Phi_2) \nonumber \\
                                        &+&\lambda_4(\Phi_1^\dagger\Phi_2)(\Phi_2^\dagger\Phi_1)+ \frac12\left[\lambda_5(\Phi_1^\dagger\Phi_2)^2 + \hc\right], \label{v12}\nonumber \\
\label{v1}
\eeqa
where the $\lambda_5$ is chosen to be real without loss of generality.

For $m_1^2 >0,~ m_2^2 <0$, only $\Phi_1$ acquires a vev after symmetry breaking.  The following scalar masses remain, 
\beqa
M^2_{\Phi^0}&=&\lambda_1/2 \nu^2, \nonumber \\
M^2_{\eta^\pm}&=&m_\eta^2+\half\lambda_1\,v^2, \nonumber \\
M^2_S&=&m_\eta^2+\half(\lambda_1+\lambda_2+\lambda_3)v^2
=M^2_{\eta^\pm}+\half(\lambda_2+\lambda_3)v^2, \nonumber \\
M^2_A&=&m_\eta^2+\half(\lambda_1+\lambda_2-\lambda_3)v^2
=M^2_{\eta^\pm}+\half(\lambda_2-\lambda_3)v^2,
\label{inmass}
\eeqa

The Yukawa Lagrangian contains
\begin{eqnarray}
{\cal L} \supset &-&-\bar l^i_L Y^e_{ij} \Phi_1 l^j_R -\bar l^4_L Y^e_{4i} \Phi_1 N_i  ~h.c. \nonumber\\
&-&\bar l^i_L Y^f_{ij} \Phi_2 N_j +~h.c.
\end{eqnarray}
And there is an additional Majorana mass term,
\begin{equation}
M_i N_iN_i.
\label{nmass}
\end{equation}
Yukawa couplings with $\Phi_1$ generates mass for fourth generation quarks and leptons while the SM neutrinos remain massless at tree level. 

Three massless neutrinos and four massive Majorana neutrinos remain after diagonalization of the $7\times7$ neutrino mass matrix.  The active neutrinos receive their Majorana mass via the one-loop diagram displayed in the model detailed in \cite{Ma:2006km}, see fig. \ref{1loop}.  The SM neutrino mass matrix is 

\begin{equation}
{\cal M}^\nu_{ij} \simeq \sum_{k=1}^3 \frac{Y^f_{\text{ik}}Y^f_{\text{kj}}M_k}{16 \pi ^{2 }}\left[\frac{\chi _3{}^2}{\chi _3{}^2-M_k{}^2}\text{ln} \frac{\chi _3{}^2}{M_k{}^2}-\frac{\eta _3{}^2}{\eta _3{}^2-M_k{}^2}\text{ln} \frac{\eta _3{}^2}{M_k{}^2}\right].
\label{nmass}
\end{equation}

$Y_{ij} \sim 10^{-2}, M_k \sim 100 \gev$ and the scalar masses of order 100 GeV gives the approximate SM neutrino masses.  

Majorana neutrino dark matter in the inert doublet model have been studied extensively. In ref. \cite{Suematsu:2011va}, the author use the two Higgs model to explain the abundance cold dark matter and leptogenesis, the lightest of exotic particles, $N_1$ is of the order of TeV scale.  In \cite{Gelmini:2009xd}, we may have a non-thermal dark matter $N_1$ to be on the order of a few GeV to MeV scale while the scalars and the other heavier singlet  fermions of the order of $\sim 100-200$ GeV. 

Presently, we have shown a model in which SM neutrinos derive naturally small neutrino masses while the fourth generation neutrino can obtain a large mass. However, it may seem strange that the heavy  fourth generation quarks and the fourth generation neutrinos obtain their large masses from the same Higgs doublet as do other SM particles. And so we would like to explore another option in which they derive their masses through Yukawa couplings with a third Higgs doublet.  

\section{Fourth generations with a third inert doublet}
Three Higgs doublet models have been studied in the past. Such models easily accommodate new sources of CP violation in the quark sector \cite{Branco:1980sz, Branco:1985pf}. They also allow for loop-induced dimension seven operators, which can generate tiny neutrino masses \cite{Kanemura:2010bq}. In our model, the SM neutrino acquire mass through loop-induced dimension five operators while the fourth generation neutrino acquire a tree-level mass. 

Here, the SM is extended with two Higgs doublets, $\Phi_1, \Phi_2, \Phi_3$, three extra gauge singlet fermions $N_i$, an extra lepton generation  $L^4_L, L^4_R$, and quarks $U_R, U_L$.   We impose a discrete ${Z_2}' \times Z_2$ symmetry to forbid flavor changing neutral currents in the SM quark sector.  
\begin{table}[ht]
  \centering
  \begin{tabular}{|c|ccc|ccc|cc|cc|cc|ccc|}
\hline
  & $Q_{Li}$& $u_{Ri}$ & $d_{Ri}$& $Q_{L}^4$& $U^4_{R}$ & $D^4_{R}$& $L_i$ & $e_{Ri}$&$L^4_L$ & $E^4_{R}$ & $N_{Ri}$ & $\nu^4_{Li}$ & $\Phi_1$ & $\Phi_2$ & $\Phi_3$  \\
\hline

$SU(2)_L$ &$2$ & $1$& $1$& $2$ & $1$& $1$&$2$ & $1$& $1$& $1$& $2$& $2$& $2$& $2$& $2$\\
$U(1)_Y$& $\frac 16$ & $ \frac 23$ & $-\frac 13$& $\frac 16$ & $ \frac 23$ & $-\frac 13$& $-\frac 12$ & $-1$ &$-\frac 12$ & $-1$ &$0$ &$0$ &$\frac12$ & $\frac12$&$\frac12$\\
\hline
$Z_{2}$& $+$ &$+$ &$+$ &$+$ &$-$ &$-$ & $+$ & $+$ &$-$ & $+$ & $+$ & $+$ & $+$&$ -$&$+$ \\
\hline
$Z_{2}'$& $+$ &$+$ &$+$ &$+$ &$+$ &$+$ & $+$ & $+$ &$+$ & $+$ & $-$ & $+$ & $+$&$ +$&$-$ \\
\hline
\end{tabular}
\end{table}

We shall follow the formalism as outlined in \cite{Grzadkowski:2010au}, The most general renormalizable scalar potential allowed by the symmetries is

\begin{equation} \label{Eq:fullpot}
V(\Phi_1,\Phi_2,\Phi_3)
= V_{12}(\Phi_1,\Phi_2) + V_3(\Phi_3) + V_{123}(\Phi_1,\Phi_2,\Phi_3)
\end{equation}
We refer to the sector containing $\Phi_1$ and $\Phi_2$ as the non-inert sector and  $\Phi_3$ as the inert doublet.  These sector potentials are 

\beqa
V_{12}(\Phi_1,\Phi_2) &=&-\frac12\left\{m_{11}^2\Phi_1^\dagger\Phi_1+ m_{22}^2\Phi_2^\dagger\Phi_2 + \left[m_{12}^2 \Phi_1^\dagger \Phi_2+ h.c\right]\right\} \nonumber \\
                                        &+&\frac{\lambda_1}{2}(\Phi_1^\dagger\Phi_1)^2+ \frac{\lambda_2}{2}(\Phi_2^\dagger\Phi_2)^2+ \lambda_3(\Phi_1^\dagger\Phi_1)(\Phi_2^\dagger\Phi_2) \nonumber \\
                                        &+&\lambda_4(\Phi_1^\dagger\Phi_2)(\Phi_2^\dagger\Phi_1)+ \frac12\left[\lambda_5(\Phi_1^\dagger\Phi_2)^2 + \hc\right], \label{v12} \\
V_3(\Phi_3) &=& m_{\Phi_3}^2\Phi_3^\dagger \Phi_3 + \frac{\lambda_{\Phi_3}}{2}
(\Phi_3^\dagger \Phi_3)^2,
\label{v3}
\eeqa
Their mutual couplings, bilinear in the $Z_2$-odd field $\Phi_3$, are given by
\beqa
V_{123}(\Phi_1,\Phi_2,\Phi_3)
&=&
\lambda_{1133} (\Phi_1^\dagger\Phi_1)(\Phi_3^\dagger \Phi_3)
+\lambda_{2233} (\Phi_2^\dagger\Phi_2)(\Phi_3^\dagger \Phi_3) \nonumber  \\
&+&\lambda_{1331}(\Phi_1^\dagger\Phi_3)(\Phi_3^\dagger\Phi_1)
+\lambda_{2332}(\Phi_2^\dagger\Phi_3)(\Phi_3^\dagger\Phi_2) \nonumber  \\
&+&\half\left[\lambda_{1313}(\Phi_1^\dagger\Phi_3)^2 +\hc \right]
+\half\left[\lambda_{2323}(\Phi_2^\dagger\Phi_3)^2 +\hc \right]
\label{v123}
\eeqa
Here, $\lambda_{1133}$, $\lambda_{2233}$, $\lambda_{1331}$ and
$\lambda_{2332}$ are real, whereas $\lambda_{1313}$ and
$\lambda_{2323},~m_{12}$ can be complex. The soft breaking symmetry $m_{12}$ term is added to prevent a massless Nambu-Goldstone boson when the discrete symmetries $Z_2$ is broken.   $\lambda_{1313},~\lambda_{2323},~m_{12}$ terms provide  extra sources of CP violation in the Higgs potential.   

We use the concept of dark democracy as introduced in \cite{Grzadkowski:2010au}, where the couplings are enormously simplified.
\beqa \label{Eq:DarkDemocracy}
\lambda_a\equiv \lambda_{1133}&=\lambda_{2233}, \nonumber \\
\lambda_b\equiv \lambda_{1331}&=\lambda_{2332}, \nonumber \\
\lambda_c\equiv \lambda_{1313}&=\lambda_{2323} \text{ (real)},
\eeqa
Defining $\lambda_L\equiv \half(\lambda_a+\lambda_b+\lambda_c)$, the scalar masses of the inert Higgs are the same as in eqn. \ref{inmass} where $\lambda_{a,b,c}$ is substituted for $\lambda_{1,2,3}$. We also define,
\begin{equation} \label{Eq:lambda_L}
\lambda_L\equiv \half(\lambda_a+\lambda_b+\lambda_c)
=\frac{m_{S_3}^2-m_{\eta_3}^2}{v^2},
\end{equation}
After spontaneous symmetry breaking,  $\Phi_1$ and $\Phi_2$ acquire a vev.  We derive the condition to ensure $\langle\Phi_3\rangle=0$ to be
\begin{equation}
m_{\eta_3}^2+\lambda_L v^2=m_{S_3}^2>0,
\end{equation}
Where $m_{\eta_3}$ is the mass of the CP-even scalar  of the inert doublet and $M_{S_3}$ is the mass of the CP-odd scalar of the inert.

Three CP-even scalars,  two CP-odd scalars and two charged scalars remain after symmetry breaking. The CP-even scalars, $ \eta_1,\eta_2$ mix with angle $\alpha$, like in the generic two Higgs doublet model case and form the physical neutral scalars, $h$ and $H$.

\begin{eqnarray}
h = \eta_1\ \rm{cos}\alpha  +\eta_2\ \rm{sin}\alpha \\
H = \eta_1\ \rm{cos}\alpha  - \eta_2\ \rm{sin}\alpha.
\end{eqnarray}
The usual $\rm{tan}\beta = \nu_2/\nu_1$ parametrizes the mixing between $\Phi_1$ and $\Phi_2$. Presumably  $\rm{tan}\beta$ is large since $\nu_2 $ should be big enough to give mass to the heavy fourth generation quarks and leptons. 

The Yukawa Lagrangian contains
\begin{eqnarray}
{\cal L} \supset &-& \bar Q^i_L Y^u_{ij}  \Phi_1 U^j_R -\bar Q^i_L Y^d_{ij}  \Phi_1 U^j_R -\bar L^i_L Y^e_{ij} \Phi_1 E^j_R + ~h.c. \nonumber\\
&-& \bar Q^4_L Y^u_{44} \Phi_2 U^4_R -\bar Q^4_L Y^d_{44} \Phi_2 U^4_R -\bar L^4_L Y^e_{44} \Phi_2 E^4_R -\bar L^4_L Y^e_{4i} \Phi_2 N^i_R + ~h.c. \nonr\\
&-&\bar L^i_L Y^f_{ij} \Phi_3 N_j +~h.c.
\end{eqnarray}

There also exists the Majorana mass for the singlet fermions as shown in eqn. \ref{nmass}

$\Phi_1$ couples only to the SM quarks and fermions, $\Phi_2$ couples only to the fourth generation and $\Phi_3$ couples the singlet fermions to the SM model fermions. The discrete symmetry forbids the coupling between SM neutrinos, $N_i$ with  $\Phi_1$ or  $\Phi_2$.  The zero vev for  $\Phi_3$ ensures that the SM neutrinos remain massless at tree-level. 

The fourth generation quarks acquire a Dirac mass $= \nu_2 Y_{44}$.  The fourth generation neutrino acquires Dirac mass, $m_D = \nu_2 \sqrt{ \sum_i Y_{4i}^2}$.  Recent unitary constraints for a SM like Higgs give bounds of tan$\beta < 6$ which gives $\nu_2  <  1$ TeV.

\begin{figure}[htb]
\centering
\includegraphics[width=2.8in]{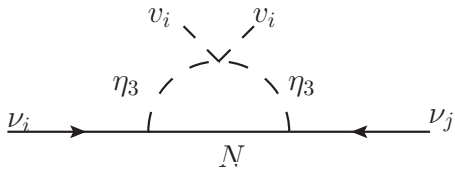}
\caption{ The 1-loop diagram which gives rise to the active neutrino masses.\label{1loop} }
\end{figure}

The SM neutrinos acquire mass the same fashion as in the two Higgs doublet case. However, two loops contribute from the $\nu_1$ and $\nu_2$ to fig. \ref{1loop}.  $Y_{ij} \sim 10^{-4}, M_k \sim 50 \gev$ and the scalar masses of order 100 GeV gives the approximate SM neutrino masses.   

Additional symmetries can be used to derive radiative mass generation through loop-induced  dimension seven operators.  \cite{Kanemura:2010bq} show several such viable mechanisms with the addition of a softly broken $Z_5$ symmetry.

The lightest stable particle charged under the discrete symmetries are good dark matter (DM) candidates. \cite{Grzadkowski:2009bt, Grzadkowski:2010au} explores scalar DM in the three Higgs doublet model without singlet fermions. They show viable candidates in the broad spectrum of masses from a few GeVs to a few hundreds of GeV.  Re-exploring the parameter space of scalar DM  with the addition of three singlet fermions is a worthwhile endeavor though outside the scope of this paper.

\section{Fourth generation and Higgs boson searches at LHC}
 As mentioned before, assuming a fourth generation model, the SM-like model Higgs is ruled out In the region $120\gev < m_H<600\gev$ at 95\% C.L. The most recent result from the LHC \cite {lhcsm3} suggests signals for the Higgs boson around$126 \gev$ in the $H \to \gamma\gamma$ channel.   In fourth generation models $H\to\gamma\gamma$ rate is suppressed even as  $gg\to H$ rate is enhanced. Fourth generation models might still be compatible \cite{He:li} if one considered signals from the $\gamma\gamma$ channel only.  However, the absence of an excess in the $WW$ channel makes this observation somewhat mute, and if we must decide on a viable model of fourth generations, it will mostly likely demand an extended scalar sector or a SM Higgs with a significant invisible width.
 
There are a variety of ways explored in the literature in which the tension between collider data and fourth generation quarks can be reduced.  Scalar color octets modifies the coupling between the quarks and the Higgs thereby reducing the production cross section \cite{He:2011ti}.  The gluon fusion rate can also be modified in the presence of vector-like fourth generation quarks \cite{Ishiwata:2011hr}.
 
The exclusion can be evaded for a light Higgs if the Higgs has a significant invisible width. For example, if the Higgs decays into a light fourth generation Majorana neutrino, \cite{Keung:2011zc} showed that Higgs can still be allowed in $120\gev < m_H<155\gev$.  A similar result also occurs in a two Higgs model with an inert doublet. In this case Higgs could decay into a light inert scalar. The authors of \cite{Melfo:2011ie} show that the $120\gev < m_H<150\gev$ is still a viable window for a fourth generation model.  The inert scalar could be DM or could decay to the light singlet fermions.

But the case of three Higgs doublets, the results of \cite{Melfo:2011ie}  are slightly modified, since there is the mixing of the CP-even scalars in the non-inert Higgs sector.  Suppose $h$ is the light of the CP-even scalar and SM-like, meaning $\alpha \sim \beta$, the production cross section of $h$ is still enhanced as in the case without extra Higgs doublets. However, its invisible decay width can be enlarged, if it decays into light inert scalars. Also the three Higgs scenario allows for a bigger invisible width than the two Higgs scenario because of the additional scalars from the third doublet. The authors of \cite{Chen:2012wz} have investigated the parameter space in which the fourth generations with two mixed Higgs doublets is compatible with recent LHC results. They have shown that fourth generations is still allowed in the $120\gev < m_H<150\gev$  region, even if one took into account hints in $H\to\gamma\gamma$ channel. 

On the other hand, suppose $h$ is the heavier Higgs boson and not SM-like. \cite{He:2011ti} notes that the unitarity bounds on a general two-Higgs model allows for $m_h \sim 700\gev$. How these unitarity bounds change with the addition of a third inert doublet  is worth investigating further.  At the very least in the non-inert sector of the scalar potential, the unitarity bounds are unchanged.

\section{Conclusions}
The model above presents a framework of naturally small SM neutrino and heavy fourth generation neutrino while allowing for a variety of DM candidates. We have also shown that these fourth generation models are compatible with the LHC Higgs searches in the low Higgs mass region, $120\gev < m_H<150\gev$ and in the high mass region  $600\gev < m_H \sim 700\gev$.

\section{Acknowledgments}
I would like to thank Sandhya Choubey for helpful discussions.
\

\end {document}